\documentstyle[12pt,draft,epsf]{article}

\newcommand \be  {\begin{equation}}
\newcommand \ee  {\end{equation}}
\newcommand \bea {\begin{eqnarray} \nonumber }
\newcommand \eea {\end{eqnarray}}

\newcommand \gs   {\sigma}

\newcommand \vs {{\bf s}}

\begin{document}

\title{The Fully Frustrated Hypercubic Model is Glassy and
       Aging at Large $D$}
\author{Enzo Marinari$^{(a)}$, Giorgio Parisi$^{(b)}$
        and Felix Ritort$^{(a,b)}$\\[0.5em]
  {\small (a): Dipartimento di Fisica and Infn, Universit\`a di Roma
    {\em Tor Vergata}}\\
  {\small \ \  Viale della Ricerca Scientifica, 00133 Roma (Italy)}\\
  {\small (b): Dipartimento di Fisica and Infn, Universit\`a di Roma
    {\em La Sapienza}}\\
  {\small \ \  P. A. Moro 2, 00185 Roma (Italy)}\\[0.5em]}
\date{October 8, 1994}
\maketitle

\begin{abstract}
We discuss the behavior of the fully frustrated hypercubic cell in the
infinite dimensional mean-field limit. In the Ising case the system
undergoes a glass transition, well described by the random orthogonal
model. Under the glass temperature aging effects show clearly.
In the $XY$ case there is no sign of a phase transition, and the system
is always a paramagnet.
\end{abstract}

\vfill

\begin{flushright}
  {\bf  cond-mat/9410089}
\end{flushright}

\newpage

%234567890123456789V123456789T123456789Q123456789C123456789S123456789S123456789
%%%%%%%%%%%%%%%%%%%%%%%%%%%%%%%%%%%%%%%%%%%%%%%%%%%%%%%%%%%%%%%%%%%%%%%%%%%%%%%

Recently there has been a renewed interest in the study of deterministic
models (without quenched disorder) with a complex low temperature behavior
\cite{MAPARA,MAPARB,BOUMEZ} (for more details about the dynamics see
\cite{MIGLIO,MIGRIT}, for additional developments see \cite{FRAHER,CKPR}).
In our first paper, ref. \cite{MAPARA}, we have discussed the low
autocorrelation model, and we have solved it by using replica theory.
In the second paper of the series, ref. \cite{MAPARB}, we have introduced
an other class of models that share the same crucial features, which are in
same sense more generic than the low autocorrelation sequences.
Let us briefly describe them. They are, as we have already said, deterministic
models, which do not contain frozen disorder. Their Hamiltonian is based on
a long range interaction

\be
  H \equiv -\sum_{x,y} J_{x,y} \gs_x \gs_y\ ,
  \label{E_HAMSC}
\ee

\noindent
where the indices $x$ and $y$ run from $1$ to the number of sites
$N$. The couplings $J_{x,y}$ are not random variables, but they are
defined as ({\em sine model})

\be
  J_{x,y}=\frac{1}{\sqrt{2N+1}}\sin \bigl ( \frac{2\pi xy}{2N+1} \bigr )\ .
\ee

We notice \cite{MAPARA,MAPARB} at first that the high-temperature
expansion of this class of models can be computed in a
straightforward way after noticing that the couplings $J_{x,y}$ form an
orthogonal matrix. By proceeding on the same lines the model defined
by the Hamiltonian (\ref{E_HAMSC}) can be solved
completely. One substitutes the orthogonal $J_{x,y}$ matrix with a
distribution of generic orthogonal random matrices, integrates over
them (we call this model the random orthogonal model, $ROM$), and
solves the model by using replica theory. The solutions of the random
model (based on generic orthogonal coupling matrices) and of the
deterministic model coincide, both in the high $T$ phase and in the
spin glass low $T$ phase (but for special properties that the
deterministic model can have for special $N$ values at low $T$).

The static and dynamical behavior of these systems can now be
understood from the point of view of the usual analysis of disordered
systems. They show two very distinct transitions.  The first one, at
$T_{RSB}$, corresponds to the usual static transition.  Here replica
symmetry breaks. At $T=T_{RSB}$ the entropy is very small\footnote{In
the context of structural glasses $T_{RSB}$ corresponds to the
Kauzmann temperature \cite{KAUZMA}.}.

The other transition appears in the solution of the dynamical
behavior. We call it the glass transition temperature $T_G$. Below
this temperature strong non-equilibrium effects start to appear. In
\cite{MAPARA} we have shown that this transition exists even in the
deterministic models we have described before.

In this note we will show that the hypercubic fully frustrated Ising
lattice model in the limit of an infinite number of dimensions
undergoes a glass transition of the same kind of the one we have found
in the low autocorrelation and the sine model.  We will show that such
a glass transition is well described by the $ROM$. We will show that
the model undergoes aging.  The glass transition exists only for
Ising-like variables. We will show that in the $XY$ case of
continuous, compact variables there is no phase transition. Here the
system is always paramagnetic, and the $T=0$ ground state is reached
smoothly when cooling from high $T$.

The Hamiltonian of the fully frustrated lattice model is

\be
  \protect\label{E_H2}
  H \equiv -\frac{1}{\sqrt{D}}\sum_{(x,y)}\, J_{x,y}\  \vs_x \vs_y\ ,
  \ \ | \vs_x | =1\ ,
\ee

\noindent
where the sites $x$ lie on a hypercubic cell in $D$ dimensions (i.e.,
the $x$ can take the values $[0,1]^D$), and the sum runs over all
nearest neighbor couples.  The $J_{x,y}$ couplings take the values
$\pm 1$, and are such that all plaquettes are
frustrated (i.e., the product of the $J$'s along the bonds of each
plaquette is $-1$).  The $\vs_x$ variables live on the $n$-dimensional
sphere, where $n$ is the number of components of the spin vectors.
$n=1$ for the Ising case and $n=2$ in the $XY$ case.

The condition of being fully frustrated can be implemented in an
infinity of ways. We have used two of them, which are gauge equivalent
(one can go from one to the other by a gauge transformation on the $J$
and the spins).  We know about the first one from Lattice Gauge
Theories, as a tool to discretize lattice fermions
\cite{KAWSMI}. $\mu$ and $\nu$ run over the lattice directions, from
$1$ to $D$, and one sets

\be
  J_{x,y}\equiv J(x_{\mu},x_{\mu}+e_{\nu})
  = (-1)^{\sum_{\mu < \nu} x_{\mu}} \ .
  \label{E_KSGAUGE}
\ee

\noindent
Here $x_{\mu}$ labels the $D$ components of the coordinates of the
site $x$, and $e_{\nu}$ is the unit vector in the direction $\nu$,
which connects the site $x$ to the site $y$. The $J$ in the direction
$1$ have the value $+1$.  The $J_{x,y}$ form an orthogonal matrix,
strengthening our expectation that the dynamical behavior of this
model will be described by the $ROM$.

We have also used (luckily enough with identical results) the
construction suggested in ref. \cite{DPTV1,DPTV2}, which can be
defined by induction. Let us take in $D=1$ all couplings equal to $1$.
Now we define how to construct a fully frustrated $D+1$ dimensional
simplex when given a $D$ dimensional one. One just has to duplicate
the $D$ dimensional simplex, multiply all coupling of the second copy
times $-1$ (i.e, flipping all links), and join the corresponding sites
of the two simplicia with $+1$ couplings. It is easy to see that this
procedure generates a fully frustrated lattice in $D+1$ dimensions.

We have also used a hypercubic lattice in $D$ dimensions. On our
lattice (a single cube) there are $2^D$ spins, and each spin is
connected to $D$ first neighboring sites.

Former work on the subject is contained in
ref. \cite{DPTV1,DPTV2,YEDGE1,YEDGE2}.  Let us first focus on the
Ising case ($n=1$). It is easy to see \cite{DPTV1,DPTV2} that a lower
bound for ground state energy $E_0$ of the model is (given the
normalization of eq. (\ref{E_H2}))

\be
  E_0 \ge -\frac12\ .
  \protect\label{E_BOUND1}
\ee

\noindent
Also one can see that this bound is independent from the number of spin
components $n$.

For Ising spins the bound (\ref{E_BOUND1}) can be improved
\cite{DPTV1} in the cases where the dimension is not a square integer.
The constraint that all spins are integer implies that
non-trivial Diophantine equalities have to been satisfied from
admissible spin configurations. The authors of \cite{DPTV1} succeed to
exhibit a class of spin configurations that saturates the improved
bound up to $D=7$. For $D \ge 8$ they cannot be sure if configurations
that satisfy their bound exist.

We have used a cooling procedure and a Monte Carlo annealing scheme
(simulating from high temperatures $T$ a thermal cycle down to $T=0$)
to gather information about the ground state structure. Let us note
at first that the simple cooling procedure is not efficient, and that
the Monte Carlo annealing is crucial to get reasonable results.

For $D$ going from $3$ to $7$ we find the same ground state exhibited
in ref. \cite{DPTV1}. In $D=6$ we confirm that only configurations
corresponding to the first solution of table $I$ of  \cite{DPTV1} are
realizable, and that the other solution of the Diophantine equation
seems not to correspond to any spin configuration. In $D=8$ we have
been able to exhibit the ground state corresponding to the improved
energy lower bound.

In $D=9$, the most accessible perfect square $D$ value after the easy
case of $4$, we have not been able to access a ground state saturating
the $-\frac12$ bound, but we have reached a value very close to that,
$-.494792$. To give the scale we notice that on a scale where the
minimal energy jump is $1$ the ground state energy is $-384$. On this
scale we have reached a value\footnote{We are $4$ elementary units far
from the improved bound value, but very probably the spin
configuration is completely different from the correct ground state.}
of $-380$, where the improved lower bound at $D=8$ gives $-361$, while
at $D=10$ it gives $-364$. We give these numbers to make clear we have
found that the case $D=9=3\cdot 3$ admits a specially deep ground
state energy valley. For $D \ge 10$ we do not converge to the ground
state. It is also interesting to notice that in the case $D=16$ we do
not succeed to go especially low in our search of the ground state
energy (the result of our search for $D=16$ is not far better than for
$D=15$). The picture one is uncovering here is very much similar to
the one of the deterministic models of \cite{MAPARA,MAPARB}, the main
difference being that here we have corrections of order $\frac1D$
(which makes the fully frustrated model closer to the low
autocorrelation model than to the sine model). There are special
values of the number of elementary bit for which the system ground
state is very low. The free energy landscape is very much golf course
like.  Deep valleys are very steep, and impossible to find in the
limit of large volume. The thermodynamical behavior of the system is
not influenced from these special minima.

After discussing the $T=0$ properties of the system we have been
investigating its finite $T$ thermodynamical properties. The methods
we have introduced in \cite{MAPARA,MAPARB} (where we address the reader
interested in the details of the computation) allow to solve the model
easily, by using the $ROM$ analogy. One starts by substituting
the interaction matrix $J_{x,y}$ (which, we noticed, happens to be an
orthogonal matrix) by a generic random orthogonal matrix. The new,
disordered model, is defined by means of the group invariant
integration over orthogonal matrices. Replica theory allows to get a
solution for this model, and replica symmetry breaking allows to
deal with the model even deep in the broken phase. In the high $T$
region, where the physical solution is replica symmetric, one finds
that the free energy density $f$ and the energy density $e$ are given
by

\bea
  \protect\label{E_EQRS}
  f &=& -\frac{1}{2\beta}G(\beta)-\frac{1}{\beta}\log(2)\ ,\\
  e &=&-\frac{G'(\beta)}{2}= -\frac{\sqrt{1+4\beta^2}-1}{4\beta}\ ,
\eea

\noindent
where the function $G(z)$ is given by

\be
  \protect\label{E_PRIMA}
  G(z)=- \frac{1}{2}\log(\frac{\sqrt{1+4z^2}+1}{2})+
  \frac{1}{2}\sqrt{1+4z^2}-\frac{1}{2}\ .
\ee

\noindent The high $T$ result coincides with the one derived in ref.
\cite{DPTV2} by means of diagrammatic expansion, and reobtained in a
different framework in \cite{YEDGE2}. The entropy of the high $T$,
replica symmetric solution, becomes negative below the replica
symmetry breaking point at temperature $T_{RSB}\simeq 0.0625$. Since
we are dealing with spin variables that can only take discrete values
that necessarily means that there has to be a phase transition close
to $T_{RSB}$. Following the strategy we discussed in
\cite{MAPARA,MAPARB} we can solve the replica equations by
implementing the marginality condition.  The method seems to work very
well to find \cite{MAPARA,MAPARB} the dynamical glassy temperature,
$T_G$. At $T_G$ the system enters a glassy phase. Applying the
marginality condition we obtain a glass transition temperature $T_G
\simeq 0.14$. Below this temperature we expect the dynamics to become
very slow. Energy relaxes to its equilibrium value on exponentially
large time scales.  In the glassy phase we expect the energy to stay
close to its value at $T_G$, $E_G\simeq-0.47$.

\begin{figure}
\epsfxsize=400pt\epsffile{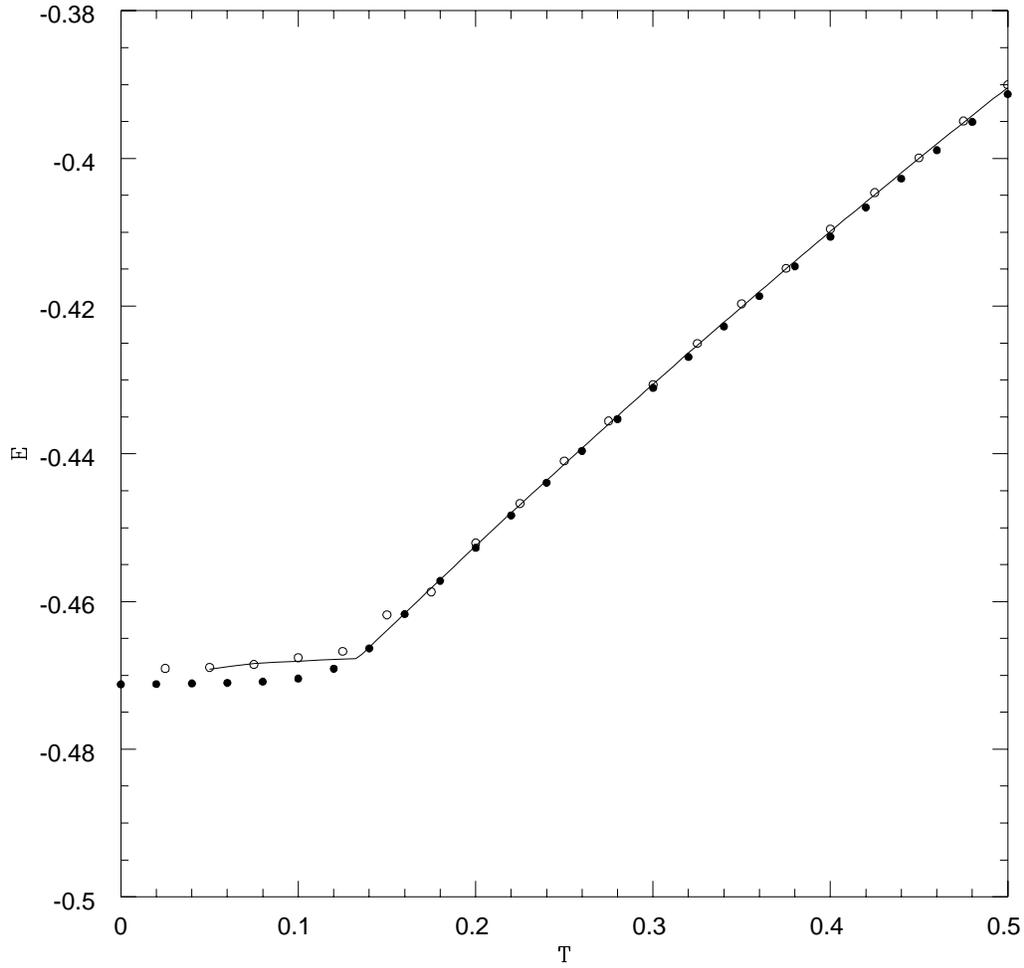}
\caption[a]{\protect\label{F_MCRUN}
Energy of the Ising fully frustrated
hypercubic cell for $D=16$ as a function
of the temperature (filled dots). With empty dots we report the results of
numerical simulations of the $ROM$ (N=186).
The  continuous line is the theoretical prediction for
the $ROM$ (in the cold phase the result is computed by assuming the
marginality condition).}
\end{figure}

We have run Monte Carlo simulations of the fully frustrated model for
different values of the dimensionality, up to $D=17$. We have been
starting our runs from a random initial configuration at $T=0.5$ and
we have slowly decreased the temperature. We show in figure
(\ref{F_MCRUN}) the energy from our Monte Carlo runs as a function of
the temperature for $D=16$, together with a numerical simulation of
the $ROM$ model.  The theoretical value in the cold phase is obtained
by using the marginality condition (see \cite{MAPARB} for details).
The results strongly fluctuate depending on the space dimensionality
(i.e., in our case depending on the volume). It is important to notice
that compared to the models we have discussed before in
\cite{MAPARA,MAPARB} the fully frustrated Ising hypercube has very
strong finite $D$ effects.  That is connected to the fact that the
energy at the glassy transition, $E_G\simeq-0.47$, is close to the
energy of the ground state for low values of $D$. Only in the regime
where $E_0(D) \ll E_G$ we expect to be able to get a clear picture of
the glass transition. This is what starts to happen for the higher
values of $D$ we have been able to study.

We discuss next in more detail the dynamical behavior of the system. We
expect that above the glass temperature $T_G$ the system behaves as a
paramagnet and time correlation functions decay very fast to zero. Below
$T_G$ aging effects appear, and the decay rate of the time correlations
depends on the history of the system. This is a common scenario
in disordered systems \cite{PDYNAM,AGING}. We have measured the spin-spin
correlation function

\be
C(t_w,t_w+t)=\frac{1}{N}\,\sum_{i=1}^N\,s_i(t_w)\,s_i(t_w+t)\  ,
\label{E_COR}
\ee

\begin{figure}
\epsfxsize=400pt\epsffile{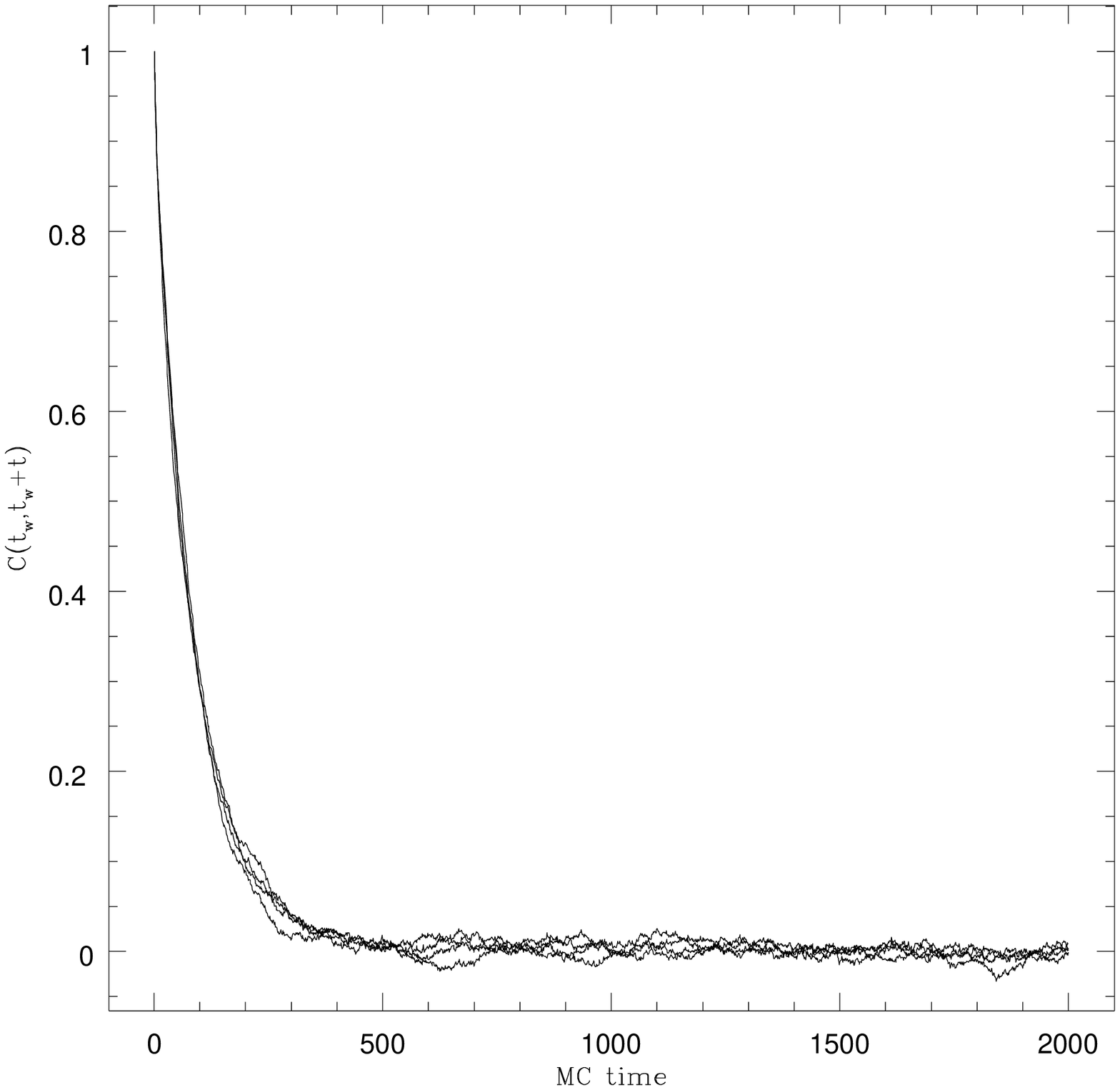}
\caption[a]{\protect\label{F_AGING20}
The correlation function $C$ for for $T=0.20$ and $D=15$.
$t_w=25$, $100$, $400$ and $1600$. Here there is no aging.}
\end{figure}

\begin{figure}
\epsfxsize=400pt\epsffile{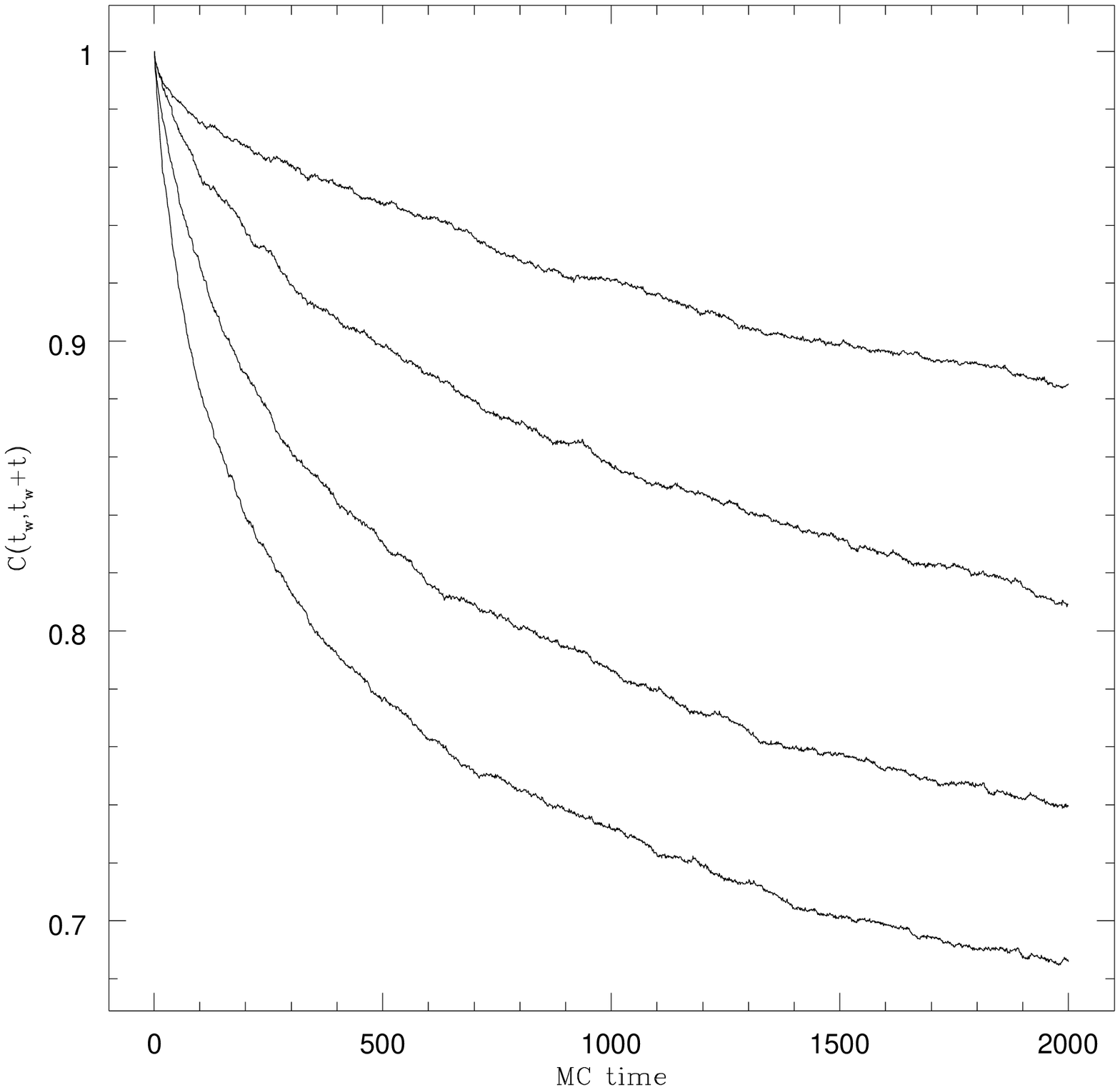}
\caption[a]{\protect\label{F_AGING10}
As in figure (\ref{F_AGING20}), but $T=0.10$. Here aging is very clear.}
\end{figure}

\noindent i.e. the correlation of the spins after a waiting time $t_w$
with the ones after the successive $t$ steps. We have observed that the
shape of the correlation function fluctuates when changing the dimension
even for very large sizes. This reflects how strong are the finite
dimensional corrections also in the metastable glassy phase. In figure
(\ref{F_AGING20}) we show the aging curves at $T=0.20$ for different waiting
times in the case $D=15$. Here we are in the paramagnetic phase, and
there is no aging. In next figure (\ref{F_AGING10})
we analyze the same correlation
functions in the broken phase, at $T=.10$. Now the aging is very clear.

\begin{figure}
\epsfxsize=400pt\epsffile{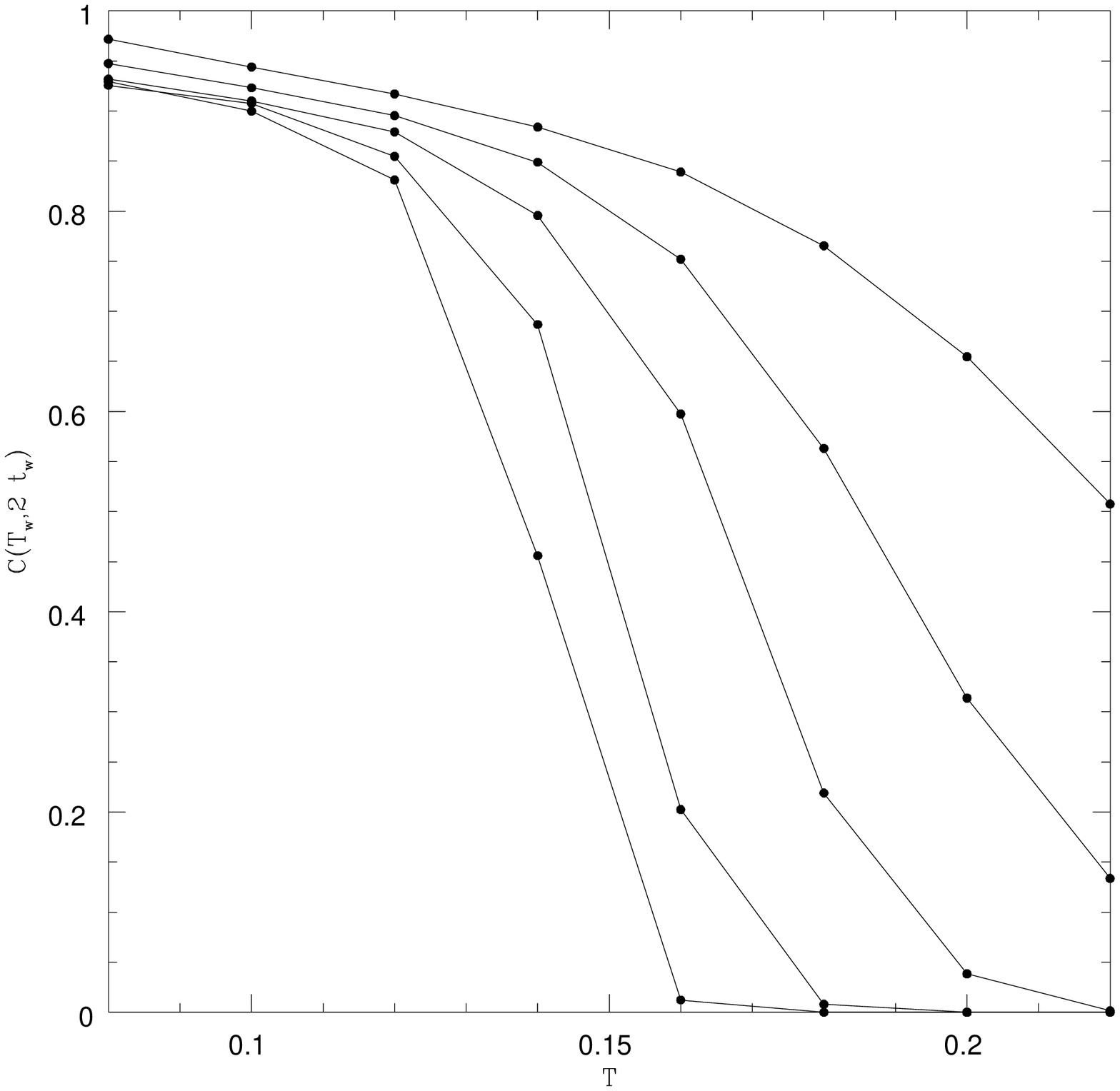}
\caption[a]{\protect\label{F_INSETTO}
$C(t_w,2t_w)$ as a function of the temperature for
different values of $t_w=30,100,300,1000,3000$ and $D=17$. Lower
curves are for higher values of $t_w$}
\end{figure}

To make explicit the discontinuous nature of the glass
transition we have measured the correlation function $C(t_w,2t_w)=q(T)$
at different temperatures. This technique has been applied recently in
the case of low autocorrelation models \cite{MIGRIT} and it seems a very good
tool to pinpoint the location of the glass transition. The results are
shown in figure (\ref{F_INSETTO}). In the limit of large values
of $t_w$ we expect $q(T)$ to be zero above $T_G$. On the contrary  $q$
experiences a discontinuous jump just below $T_G$.

\begin{figure}
\epsfxsize=400pt\epsffile{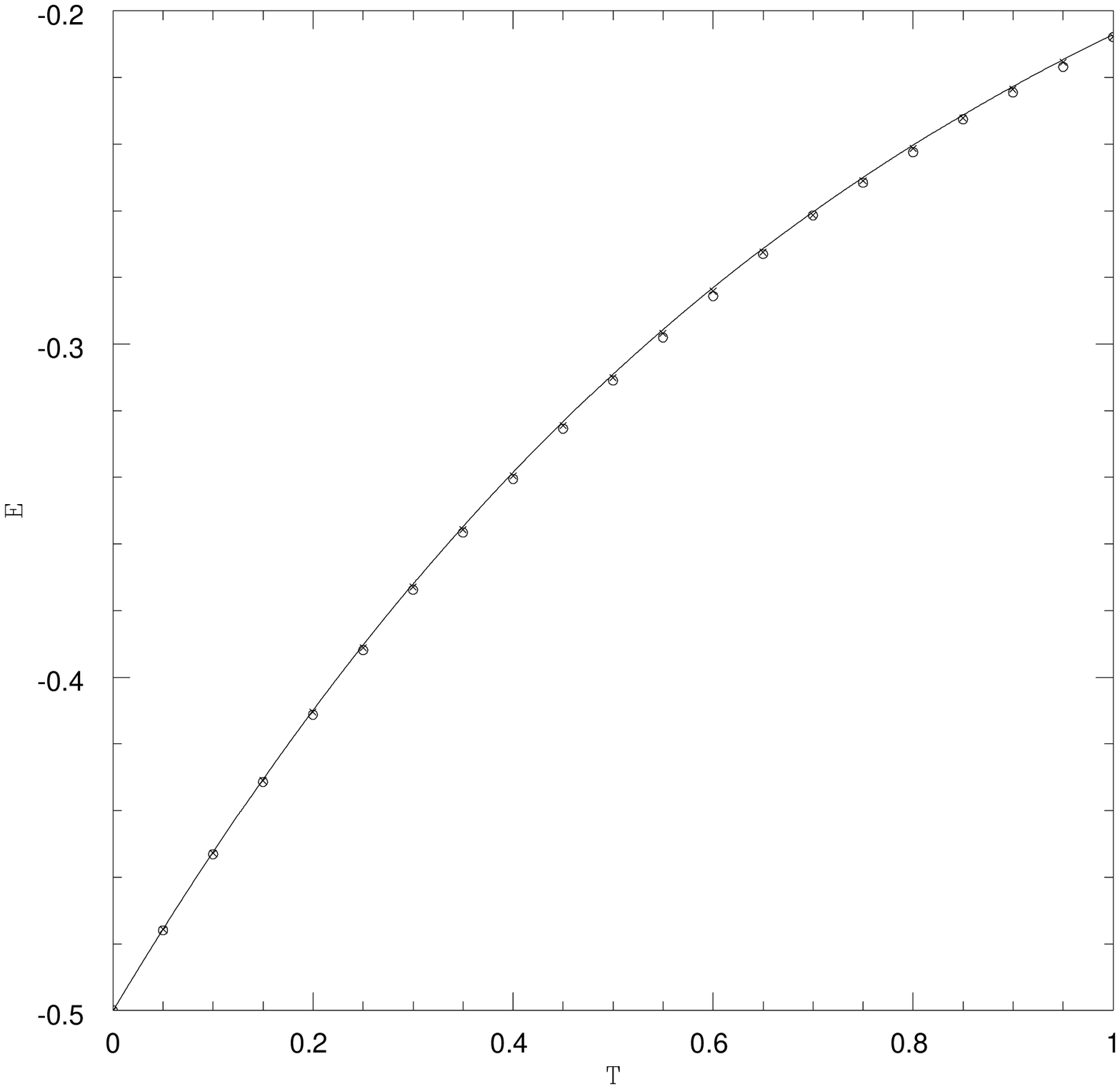}
\caption[a]{\protect\label{F_XY}
Energy of the fully frustrated hypercubic cell ($XY$ model) for
$D=8$ (empty dots) and $D=15$ (small $\times$ symbols)
as a function of the temperature $T$. The
continuous line is the high temperature
prediction. There is no signal of the presence of a phase transition.}
\end{figure}

The body of the results we have discussed for the fully frustrated Ising
model supports the standard scenario derived from the study of the $ROM$.
Now we present our results obtained for the $XY$ model, where the spin
variables are complex numbers constrained to be of modulo $1$ (and the
$J$ variables are the same as before). Here the system seems
always to be able to find its exact ground state (with $E_0=-\frac12$),
and  we have not found any evidence of the existence of a phase
transition.  The whole phase diagram in $T$ is well described by the
high-temperature expression of eq.(\ref{E_EQRS}) (after normalizing the
temperature by the number $n$ of spin components, $n=2$ in the $XY$
case). We show our results in figure  (\ref{F_XY}), where we plot the
internal energy of the model together with the high-temperature result.
In the whole $T$ region finite dimensional corrections are negligible, in
agreement with the fact that the phase space structure is very simple. We
expect similar conclusions to be valid in case of Heisenberg or spherical
spins. As far as we can understand from these results, frustration
without quenched disorder needs to be helped from the discrete nature of
the spin variables in order to create traps dangerous enough to enforce
a complex behavior.

We have shown that the Ising fully-frustrated hypercubic cell in the
high $D$ mean-field limit displays a low $T$ glassy behavior. This
model has a glass transition of a discontinuous type, well described
by the solution of the random orthogonal model. We have noticed that
the dynamical low temperature behavior is very sensitive to the finite
$D$ corrections. We have also investigated the $XY$ model. Here there
is no sign of a phase transition and the system is always
paramagnetic. It would be very interesting to understand if such a
glassy behavior is shared by usual short-ranged frustrated
non-disordered models.

\subsection*{Acknowledgments}
We gratefully thank Mark Potters for interesting discussions and a
critical reading of the manuscript.

\end{document}